\documentclass[mathleft
]{an}
\usepackage{amssymb}
\usepackage{graphicx}
\usepackage{times}
\usepackage{natbib}
\bibpunct{(}{)}{;}{a}{}{,}
\newcommand{\ie}{\textit{i.e.}~}
\newcommand{\eg}{\textit{e.g.}~}

\begin{document}

\Pagespan{1}{6}
\Yearpublication{0000}%
\Yearsubmission{2013}%
\Month{11}%
\Volume{000}%
\Issue{00}%

\title{Magnetic Field Instabilities in Neutron Stars}

\author{Riccardo Ciolfi
\thanks{
  \email{riccardo.ciolfi@aei.mpg.de}}
}
\titlerunning{Magnetic Field Instabilities in Neutron Stars}
\authorrunning{R. Ciolfi}
\institute{
Max-Planck-Institut f\"{u}r Gravitationsphysik (Albert-Einstein-Institut), Am M\"{u}hlenberg 1, 14476 Potsdam, Germany
}
\received{2013 Dec 20}
\accepted{2014 Jan 14}
\publonline{2014 Mar 16}

\abstract{Magnetic fields represent a crucial aspect of the physics
and astrophysics of neutron stars. 
Despite its great relevance, the internal magnetic field configuration
of neutron stars is very poorly constrained by the observations, and 
understanding its properties is a long-standing theoretical
challenge.     
The investigation on the subject is focused on the search for those
magnetic field geometries which are stable on several Alfv\`en
timescales, thus constituting a viable description of neutron star
interiors.    
Assesing the stability of a given magnetic field geometry is therefore    
an important part of this research. So far only simple
configurations, such as the purely poloidal or 
purely toroidal ones, have been studied in detail in perturbation
theory and, most recently, by means of nonlinear 
magnetohydrodynamic simulations. 
Here we review the basic results of the state-of-the-art general
relativistic nonlinear studies, discussing the present status of the
field and its future directions.  
}


\maketitle

\section{Introduction}\label{intro}

Neutron stars (NSs) are endowed with extremely strong long-lived 
magnetic fields, reaching surface (polar) strengths of $10^{13}$ G for
ordinary NSs (including the most common radio pulsars) and $10^{15}$ G 
for highly magnetized NSs, or {\it magnetars}
\citep{DuncanThomp92,Mereghetti2008}, as mainly inferred via the
simple dipole formula. Internal fields might be even
stronger, possibly by one order of magnitude or more. 
Magnetic fields of such intensity affect the physical properties of
NSs and play a fundamental role in the processes through which they 
are currently observed, from dipole radiation to magnetar flares.
Moreover, the magnetically-induced quadrupolar deformations of NSs
make them an interesting source of gravitational waves
\citep{Bonazzola1996,Cutler2002}, potenitally detectable in the near
future.

The amount of magnetic energy stored inside a NS 
and the internal magnetic field geometry are known to represent key
elements in determining its evolutionary path and emission properties.       
Recent results of magneto-thermal evolution studies
(\citealt{Vigano2013} and references therein) confirm this idea,
showing how the evolution of a magnetized NS may proceed on very
different timescales and with distinct features (\eg the presence or
absence of a bursting activity), depending on the internal magnetic
energy and distribution of electric currents. 
Despite many aspects of our present modelling and understanding of NSs
depend directly on the internal magnetic field energy and geometry
(including the gravitational wave emission mechanism mentioned above),
this feature is not constrained by direct observations, which only provide
information on the external magnetic field.  
This obviously constitutes a very strong limitation, justifying the
growing effort devoted to build realistic models decribing the
possible magnetic field configurations realized in a NS.

A widely accepted scenario for the evolution of a magnetized NS
pictures the following stages: (i) the newly-born NS is initially a
highly convective and differentially rotating hot fluid; (ii) in a very
short timescale (of the order of seconds to minutes) convection and  
differential rotation are damped, the star cools down and
rearranges towards a magnetohydrodynamic (MHD) equilibrium;
temperature keeps decreasing and within few hours it reaches the
critical values for the formation of a solid crust and the onset of
superfluidity, both in the range $10^{9}-10^{10}$ K; (iii) the
following evolution proceeds on much longer dissipative timescales
($10^{3}-10^{6}$ years).  
In the intermediate stage, while the star is still completely fluid
and well described as a non-superfluid ideal (electric) conductor, 
the magnetic field evolves on the Alfv\`en timescale, which typically
lies in the range $\tau_A \sim 0.01-10$ s depending on the magnetic   
field strength. Therefore, there is ample time for the magnetized
fluid to reach a stable MHD equilibrium or, alternatively, to lose
almost all of its magnetic energy (\eg in electromagnetic emission)
before the crust and/or superfluidity appear.      
According to the above scenario, the observation of long-lived
magnetic fields represents an evidence that, in this intermediate
stage, a configuration has been reached which is stable on Alfv\`en
timescales.    
A natural apporach to constrain the properties of the internal
magnetic field of NSs is then to (i) consider the widest range   
of possible equilibrium configurations in an ideally conducting fluid 
NS (\ie in the conditions met during the ``pre-crust'' stage) 
which are compatible with the observations and (ii) to assess their
stabilty on Alfv\`en timescales. 
In case stable configurations are found these would represent a viable  
description of a magnetized NS at the time of crust formation, 
setting realistic initial conditions for long-term magneto-thermal
evolution studies, where the dissipative processes are taken into
account and the effects of superfluidity/superconductivity are
included. Since in the long-term evolution the main features of the
internal magnetic field (\eg the distribution of energy in toroidal
and poloidal components, the multipolar structure and the overall
energy/strength) are modified very slowly, the intial configuration   
likely represent also a good qualitative description of the internal
magnetic field in the relatively young NSs observed. 

Since the early work of \citet{Chandrasekhar1953} a number or
analytical and numerical studies have been devoted to the construction 
of equilibrium models of magnetized NSs, under the common assumpitions 
of ideal MHD and pure-fluid matter, well-suited to describe the
conditions occurring before the formation of a solid crust and the
onset of superfluidity.  
These models focused at first on the simple purely poloidal and purely 
toroidal magnetic field geometries. However, already from the early
analytical work on nonrotating magnetized stars
\citep{Markey1973,Tayler1973,Wright1973} there has been growing
evidence that these simple geometries would suffer from the so-called 
Tayler (or kink) instability, acting on Alfv\`en timescales. The
unstable nature of purely poloidal and purely toroidal geometries   
was confirmed more recently in Newtonian numerical simulations
in the linear regime \citep{Lander:2011b,Lander:2011a} and, for
main-sequence stars, via nonlinear simulations
\citep{Braithwaite2006,Braithwaite2007}.   
Only in the very last years, the same system was studied for the
first time by means of nonlinear MHD simulations in general
relativity \citep{Kiuchi2008,Kiuchi2011,Lasky2011,Lasky2012,
Ciolfi2011,Ciolfi2012}, further verifying the presence of a
hydrodynamic instability in the purely poloidal and purely toroidal 
cases and providing important indications on the subsequent
nonlinear rearrangement of magnetic fields.   
In this paper (Section~\ref{stable-field}) we discuss in some detail
the setup and basic results of these simulations, which represent the
state-of-the-art of the nonlinear investigation on the stability of
magnetic field configurations in NSs.   

All the above stability studies converged to the idea that any
long-lived magnetic field configuration in a NS has to consist of
a mixture of poloidal and toroidal field components. Among the
mixed-field configurations it is worth mentioning the {\it
  twisted-torus} geometry, which recently emerged as a good candidate 
for NS interiors.  
It consists of an axisymmetric field where the poloidal component
extends throughout the entire star and to the exterior, while the
toroidal one is confined inside the star, in the torus-shaped region
where the poloidal field lines are closed (see \eg Fig.~1,2 of
\citealt{Ciolfi2013}).   
Important indications are in favour of the twisted-torus configuration
(see discussion in \citealt{Ciolfi2013}), including the results of
Newtonian simulations performed by \citet{BraithwaiteNordlund}, 
where this geometry emerged as the final outcome of the evolution of
initial random fields in a nonrotating fluid star. Those simulations
where adapted to study main-sequence stars, while the equivalent
evidence in a NS and in general relativity is still
missing; nevertheless, the results triggered a growing interest in
twisted-torus geometries, which were recently considered in
several equilibrium models of magnetized NSs, both in Newtonian
\citep{Tomimura2005,Yoshida2006,Lander2009,Lander2012,
Glampedakis2012,Fujisawa2012} and general relativistic frameworks 
\citep{Ciolfi2009,Ciolfi2010}.  
All the proposed equilibrium solutions, however, found a common
limitation to poloidal-dominated geometries, with a magnetic energy in 
the toroidal component always $\lesssim10\%$, which is in contrast
with the general expectation of a higher toroidal field contribution
(see \citealt{Ciolfi2013}); moreover, there are already indications
that poloidal-dominated configurations are unstable on Alfv\`en
timescales \citep{Braithwaite2009,Lander2012}.      
A solution to this problem has been offered most recently in
\citet{Ciolfi2013}, where it is shown how a different prescription for 
the electric currents allows to expand the space of known solutions to 
a much higher toroidal field content (without invoking surface
discontinuities in the magnetic field, as in \citealt{Fujisawa2013}).  
Twisted-torus configurations with a high toroidal magnetic energy 
content (\ie $>10\%$) are certainly among the most promising
candidates for stability. Future nonlinear studies, so far limited to
simple purely poloidal/toroidal geometries, will possibly provide the
missing evidence.  
  
\begin{figure*} 
\centering
\hskip 0.6cm
\includegraphics[width=7.cm]{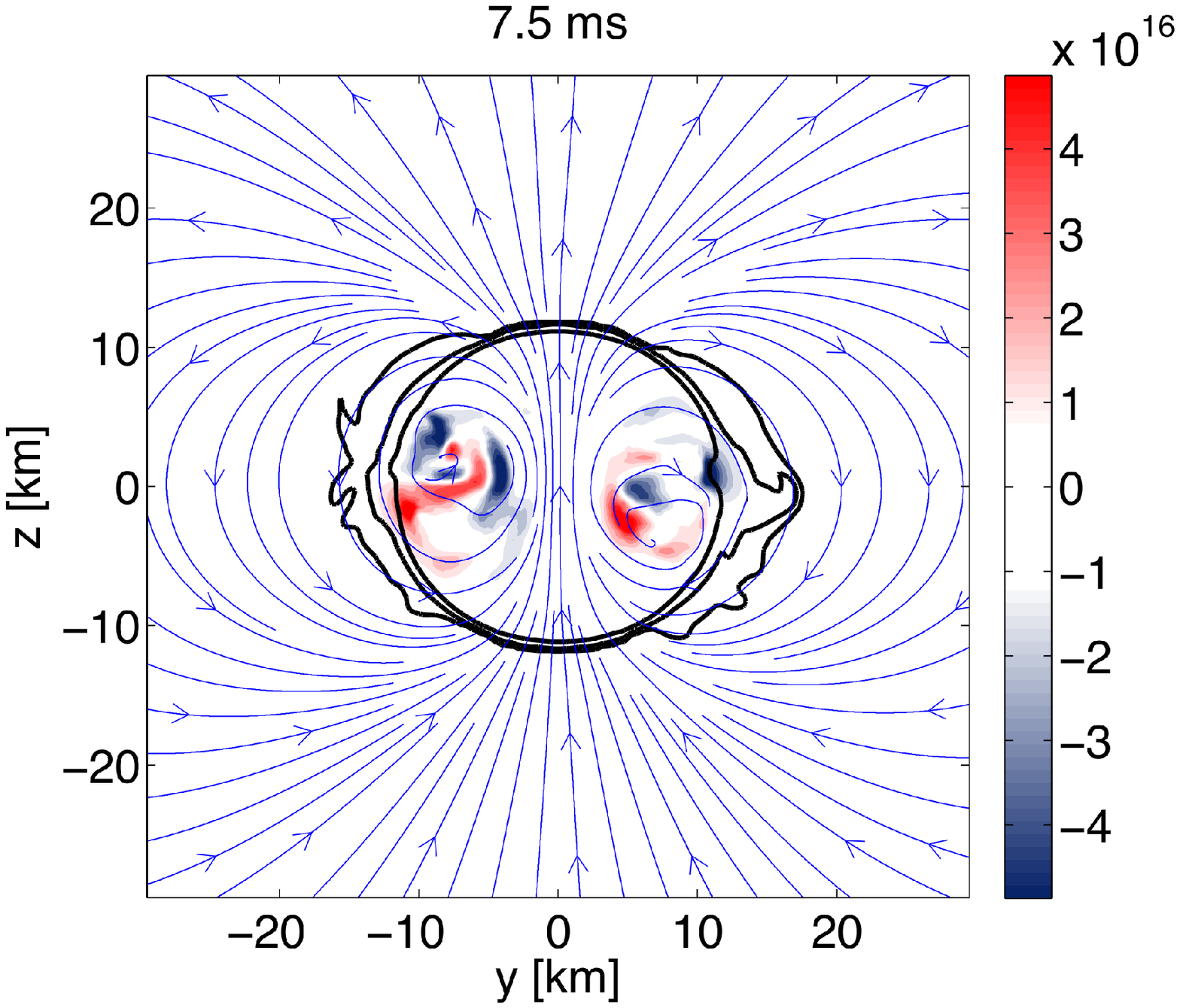} 
\hskip 0.5cm
\includegraphics[width=7.cm]{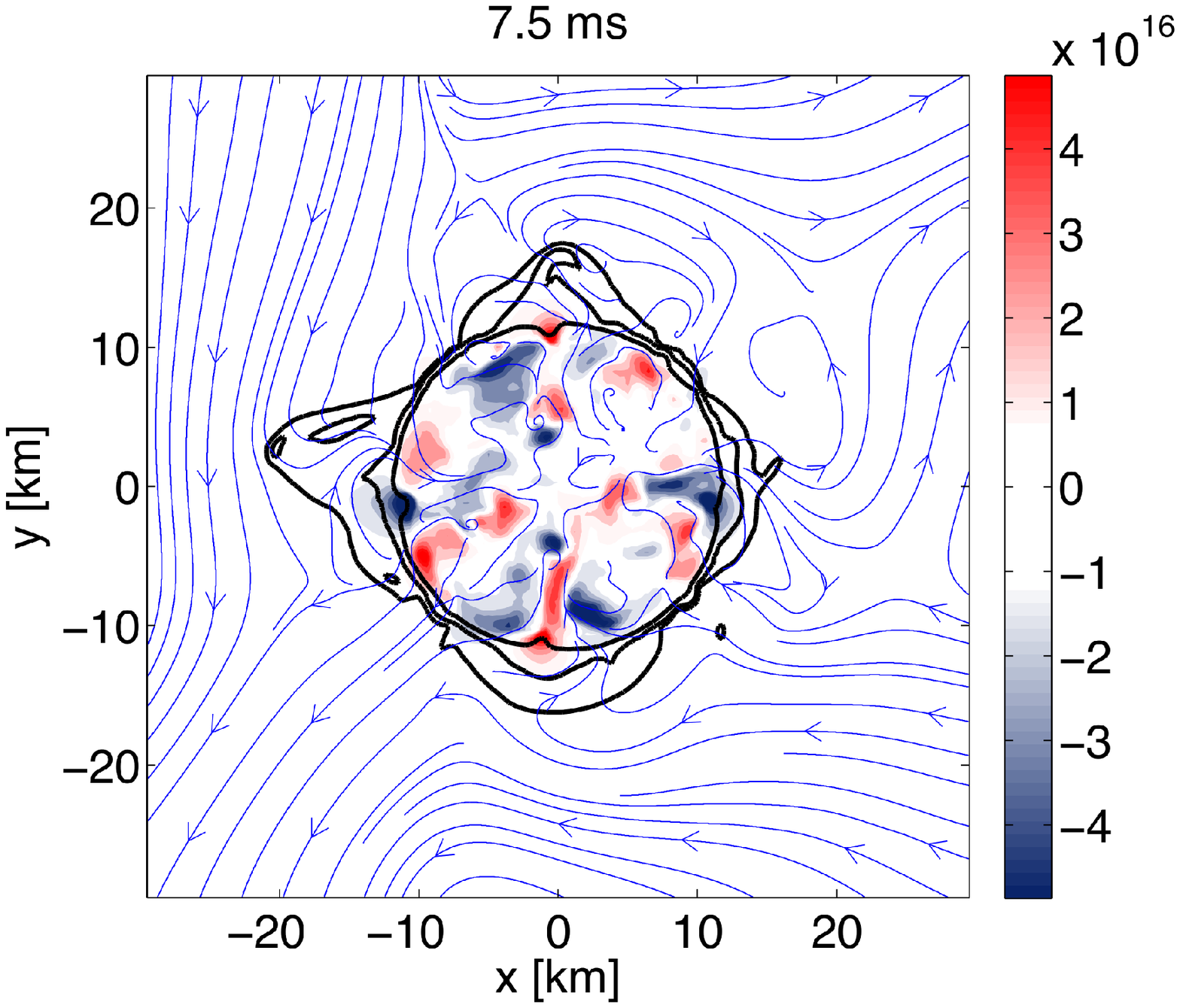}  
\caption{Instability of a purely poloidal field in a magnetized NS
  with an initial polar magnetic field strength of $6.5\times
  10^{16}$~G. Left and right panels show
  respectively the meridional 
  and equatorial view of the system at $t=7.5$~ms,
  time at which the instability has fully developed. Vector lines
  illustrate the (global) magnetic field lines, while the colors 
  show the intensity (in Gauss) of the toroidal magnetic field only;
  also reported are three rest-mass isodensity contours near the
  stellar surface, corresponding to $(0.02, 0.2, 2) \times
  10^{13}\,{\rm g/cm}^3$.}   
\label{poloidal_inst} 
\end{figure*}

Apart from the search for long-lived magnetic field configurations in
NSs, there is a second main motivation for studying the nonlinear MHD
evolution of a magnetized NS.  
The global rearrangement of magnetic fields induced by a hydromagnetic
instability (as the one affecting the purely poloidal/toroidal
configurations) is a violent, strongly 
dynamical process, and soon after the magnetar model was proposed
\citep{DuncanThomp92} it was suggested as a trigger mechanism for the
giant flares of magnetars
\citep{ThompsonDuncan1995,ThompsonDuncan2001}. This ``internal 
rearrangement scenario'' still represents one of the two leading 
models to explain the phenomenology observed in magnetar giant flares,
the other one involving a large-scale rearrangement of magnetic fields
in the magnetosphere surrounding the star
\citep{Lyutikov2003,Lyutikov2006,Gill2010}.  
Investigating the dynamics associated with hydromagnetic instabilities 
in NSs can provide important constraints on the electromagnetic and 
gravitational wave emissions to be expected, according to the internal
rearrangement scenario, in coincidence with a giant flare event.       
Moreover, by comparing with the electromagnetic luminosities and 
emission timescales measured in the three giant flares already
observed, nonlinear simulations can help establish the viability of a
hydromagnetic instability as a trigger mechanism.  
In Section~\ref{GFs} we discuss the results
obtained in this direction in the recent nonlinear general
relativistic MHD studies of the instability of purely poloidal
magnetic fields in NSs
\citep{Ciolfi2011,Ciolfi2012,Lasky2012,Zink2012}. Finally, in
Section~\ref{concl} we give our conclusions.  

\section{General relativistic MHD simulations of the Tayler
  instability}\label{stable-field}  

In this Section we discuss in more detail the recent studies on the
stability of magnetized NSs via nonlinear general relativistic MHD
simulations. So far, this investigation only covered the simple purely 
poloidal and purely toroidal configurations, confirming the presence
of the Tayler instability. 

The first three-dimensional general relativistic MHD simulations of
the poloidal field instability in NSs were presented in two parallel
works \citep{Lasky2011,Ciolfi2011}, both followed by a more extended
investigation \citep{Lasky2012,Ciolfi2012}.    
These studies employ analogous initial data, produced with the
multi-domain spectral-method code \texttt{LORENE} \citep{Bocquet1995}:   
an axisymmetric magnetizd NS with a purely poloidal magnetic 
field, composed of a barotropic fluid obeying a simple
polytropic equation of state with index $n=1$, and having a typical
mass of $1.4$~M$_\odot$ (or slightly smaller) in the unmagnetized
limit. 
The magnetic field strength at the pole varies by about one order of 
magnitude around a fiducial value of $\sim 10^{16}$~G. 
Such high magnetic field strength is chosen to shorten the evolution
timescale of the system, making the simulations computationally
feasible. Even if this exceeds by far the measured values even for
magnetars, most of the results can be extrapolated back to smaller
(and more realistic) magnetic field strengths.   
Despite the differences in the numerical codes employed, the
numerical setup (\eg size of the computational domain and
boundary conditions) and the treatment of magnetic fields outside 
the star, the initial development of the instability
shows perfect agreement between the findings of
\citet{Ciolfi2011,Ciolfi2012} and of
\citet{Lasky2011,Lasky2012}. Relevant differences 
emerge later in the evolution, as the nonlinear rearrangement of the
magnetic field proceeds.  
In what follows we focus on the results of
\citet{Ciolfi2011} and \citet{Ciolfi2012}, pointing out similarities
and differences with \citet{Lasky2011,Lasky2012}. 

\begin{figure} 
\centering
\includegraphics[width=7.cm]{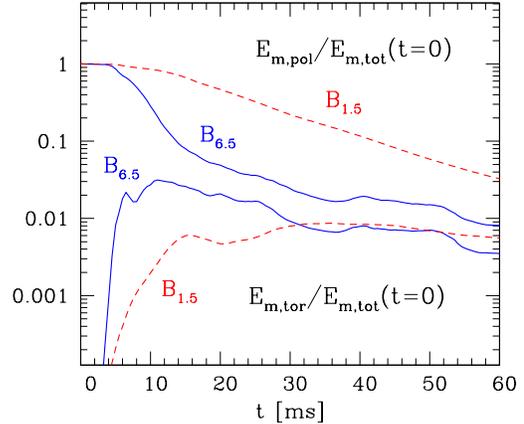} 
\caption{Evolution of poloidal and toroidal magnetic energies
  normalized to the initial total magnetic energy, in log scale. The
  continuous blue line refers to the simulation with an initial polar
  field strength of $6.5\times 10^{16}$~G, the dashed red line to the
  one with $1.5\times 10^{16}$~G.  
} 
\label{energies} 
\end{figure}

Fig.~\ref{poloidal_inst} shows two snapshots of the evolution for a 
representative simulation among those presented in
\citet{Ciolfi2012}. In this case the initial polar magnetic field
strength is $6.5\times 10^{16}$~G. 
Left and right panels give respectively the meridional and equatorial
view of the system at 7.5~ms, corresponding to the most violent phase 
of the evolution, shortly after the instability has fully developed
and the nonlinear rearrangement of the magnetized NS has begun.  
As predicted by previous linear studies, the Tayler instability first
occurs in the surrounding of the neutral line (in the closed-line
region) and forces the field to produce there a toroidal component.   
In Fig.~\ref{energies} we report the evolution of poloidal and
toroidal magnetic energies for the same simulation and for one with a 
smaller initial magnetic field strength. The different stages of the
evolution are clearly distinguishable:
(i) initially, the field is purely poloidal, but as the instability
takes place the toroidal component starts to grow exponentially; 
(ii) when a comparable local strength has been reached (around one 
Alfv\`en crossing time) the instability saturates and the nonlinear  
evolution begins; the poloidal field energy drops violently while the
toroidal energy experiences a smaller variation, resulting in a
growing toroidal-to-poloidal energy ratio; (iii) after the first
violent rearrangement, the system keeps evolving on a much longer 
timescale. 
It is worth stressing that the initial location of the instability,
the production of a toroidal component and the saturation timescales
are in full agreement with the results of \citet{Lasky2011,Lasky2012},
also meeting all the expectations from previous analytic and numerical
studies.   

The advantage of nonlinear simulations is that, in addition to
providing a confirmation to the predictions of the linear analysis, 
they allow to follow the evolution of the system beyond the
instability saturation, giving hints on the preferred state of
the system. 
The end result of the (relatively long) simulations performed in 
\citet{Ciolfi2012} gave no evidence for a stable magnetic field
configuration (although we cannot exclude that a stable condition
would be reached on much longer timescales).     
Nevertheless, some important indications were obtained.  
First, the system revealed a clear tendency to migrate towards 
a configuration where the magnetic energy is equally distributed in
the poloidal and toroidal components. This result was also reported in  
\citet{Lasky2012}. 
Secondly, the total magnetic helicity of the system, which is zero by
definition at the beginning, was always found to grow up to
significant values (see discussion in \citealt{Ciolfi2012}),
suggesting that this quantity might play a crucial role in stabilizing
the magnetic field.   
The natural conclusion is that both the equipartition of magnetic
energy in poloidal and toroidal fields and a significant amount of
magnetic helicity are likely features of any stable configuration.    

The strong loss of magnetic energy that the magnetized NS experiences
at the beginning of the nonlinear phase of the evolution, right after
the exponential growth of the instability has saturated, is mainly
associated with the diffusion of magnetic fields outside the NS.
The internal dynamics imposes 
a rapid change of the magnetic field at the stellar surface and, as a 
consequence, a reconfiguration of the external field. In a realistic
scenario, this process would load the magnetosphere with magnetic
energy and ultimately cause a strong electromagnetic emission.   
In \citet{Ciolfi2011} and \citet{Ciolfi2012}, in order to mimic this
external behaviour and provide more suitable boundary conditions for
the internal evolution, a resistive term is added in the induction
equation, which tends to dissipate the non-zero laplacian components
of the magnetic field in the exterior (\ie those which would be
rapidly carried away in electromagnetic waves according to the
Maxwell's equations in vacuum) and to restore the potential-field
condition. 
This approach, also adopted in the simulations of
\citet{BraithwaiteNordlund}, results in a more 
realistic internal dynamics, even if it represents a too crude
approximation to describe correctly the dynamics of the external
fields. Moreover, it allows for an
order-of-magnitude estimate of the amount of magnetic energy that the
star would lose during the evolution and that would be supplied to the
magnetosphere and/or to power the electromagnetic emission 
(Section~\ref{eme}).\footnote{A different apporach is adopted in
  \citet{Lasky2011,Lasky2012}, where the magnetic fields are evolved
  according to ideal MHD also in the exterior. This does not allow to
  estimate the losses of the system and is at the origin of the
  differences in the late evolution of the system.} 
Note that a small fraction of the energy lost is instead converted
into oscillatory motions, mainly damped through the emission of 
gravitational waves (Section~\ref{gwe}).

General relativistic nonlinear MHD simulations also confirmed the  
instability of NSs endowed with a purely toroidal magnetic field
\citep{Kiuchi2008,Kiuchi2011}, as predicted by linear studies.
In this case the Tayler instability acts in an anoalogous way, but
along the magnetic axis, and other instabilities may play an important  
role, \eg the Parker instability \citep{Parker1955,Parker1966}. 
The setup of these simulations is very similar to the one
employed for the purely poloidal field case. 
Also in this case, the end state of the evolutions showed no 
evidence of stable configurations. 

Rapid (uniform) rotation of the magnetized NSs has long been
suggested to have a stabilizing effect, potentially able to suppress   
the instability of purely poloidal and purely toroidal fields. 
Some of the nonlinear studies discussed in the present Section
included also the NS rotation, never finding it sufficient to avoid
the hydromagnetic instability \citep{Lasky2012,Kiuchi2011}.  

\section{Hydromagnetic instabilities and magnetar giant
  flares}\label{GFs}   

As discussed in Section \ref{intro}, nonlinear MHD simulations of
magnetized NSs subject to hydromagnetic instabilities 
can be used to test the internal rearrangement scenario of magnetar
giant flares, according to which this kind of instabilities would be
the trigger of the giant flare events 
\citep{ThompsonDuncan1995,ThompsonDuncan2001}. 
Here we discuss the first indications provided by the recent nonlinear 
studies on the electromagnetic and gravitational wave
emission that would be expected in coincidence with a giant flare.       

\subsection{Electromagnetic emission}\label{eme}

\begin{figure} 
\centering
\hspace{0.3cm}
\includegraphics[width=6.8cm]{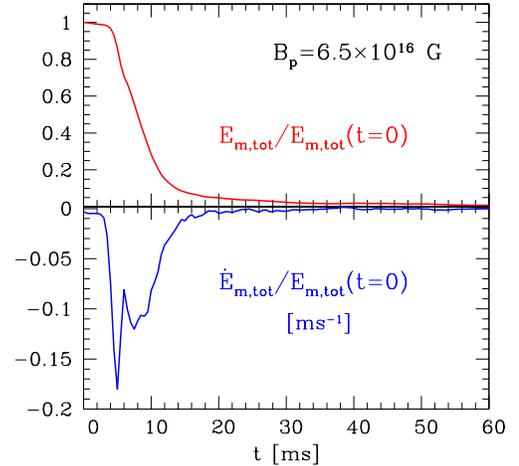} 
\caption{Top: Evolution of the total magnetic energy normalized to its
  initial value, for a NS with a purely poloidal magnetic
  field having an initial polar strength of
  $6.5\times 10^{16}$~G. Bottom: Time derivative of the total magnetic
  energy. 
} 
\label{Lem} 
\end{figure}

In this Section we report on results presented in \citet{Ciolfi2012}.
For any of the magnetic field strengths considered, the NS
always loses most of its magnetic energy at the end of the
simulations, as a result of the poloidal field instability.   
In particular, about 90\% of this energy is lost in the first few
Alfv\`en timescales after the instability saturation. 
Fig.~\ref{Lem} shows, as an example, the evolution of the total
magnetic energy $E_m$ for the same simulation shown in
Fig.~\ref{poloidal_inst}. The initial rapid drop in energy
corresponds to a spike in its time derivative, $\dot{E}_m$. 
As stated in the previous Section (\ref{stable-field}), only a small
fraction of this energy is converted into oscillatory motions,
while most of it is lost due to the diffusion of magnetic fields
outside the star. Within the approximation of the apporach adopted,
these losses give an order-of-magnitude estimate of the magnetic
energy that would leave the star because of the instability-induced
internal dynamics, supplying the magnetosphere and ultimately powering 
a strong electromagnetic emission.   

The emission observed in giant flares is characterized by an initial
spike in luminosity lasting $\sim$0.2$-$0.5~s, in which a large
fraction of the energy budget of the event is released (up to more
than 99\%), followed by a  pulsating tail lasting hundreds of seconds
and most certainly associated with the residual energy stored in the
excited magnetosphere \citep{Mereghetti2008}. 
We now assume that the trigger of these events is an internal
hydromagnetic instability and that the spike in luminosity is the
direct result of the initial violent energy release from the star,
with no significant delay due to the processing of such energy in the
magnetosphere.    
In this case, a compatibility in timescales is expected between the
observed spike emission and the typical timescale of an internal
rearrangement. 
By taking the instability of a purely poloidal field as a test case,
a comparison can be attempted between the duration of the observed
initial spike of giant flares and the duration of the spike in
$\dot{E}_m$ produced by the instability, as estimated from numerical
simulations.    
This was tried in \citet{Ciolfi2012}. After extrapolating the
results to magnetar-like field strengths the compatibility
of the two timescales was confirmed, thus providing support to
the proposed scenario.    

For the alternative, external scenario, according to which the giant
flare would 
be triggered by a reconfiguration of the external magnetic fields in
the magnetosphere, an initial spike lasting $\sim$0.2$-$0.5~s
represents instead a strong challenge, because the evolution timescale
outside the star is orders of magnitude shorter. 
The measured raise time of the spike emission,
on the other hand, is way too short to be explained with the internal
dynamical timescales, suggesting the following possible conclusion: 
the event is likely triggered by an internal rearrangement, which
provides most of the energy powering the flare, but the internal
dynamics immediately triggers also some rearrangement of the external  
magnetic fields. At this stage, however, any conclusion remains still a
matter of debate and further investigation is necessary to shed light
on the subject.  

As a note of caution, we remark that our simulations do not account  
for the presence of a solid crust, most probably relevant in the
dynamics of giant flares. Moreover, the instability of a purely
poloidal field is likely more dramatic than in a more realistic
situation, where the magnetic field would migrate from one
configuration to another with a smaller jump in magnetic energy.
\begin{figure} 
\centering
\includegraphics[width=7.28cm]{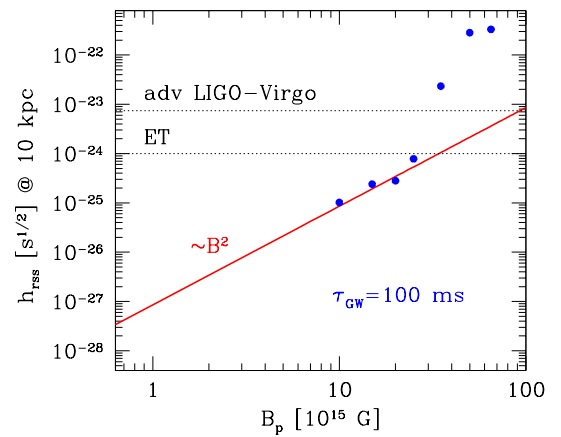} 
\caption{Gravitational wave signal amplitude versus the initial polar
  magnetic field strength, assuming a distance of 10 kpc and a damping
  time of 100 ms. Horizontal lines mark the strain-noise amplitude of
  near-future GW detectors at the f-mode frequency. The red line
  is obtained by imposing a quadratic scaling and fitting the
  results for the lowest magnetic fields (first four points). 
} 
\label{gw} 
\end{figure}
\subsection{Gravitational wave emission}\label{gwe}

The violent reorganization of magnetic fields induced by a
hydromagnetic instability is accompanied by a significant excitation
of NS oscillations, in particular in the f-mode, which can then lead
to a strong emission of gravitational waves (GWs). 
Following the argument that the instability of a purely poloidal field
represents a test case for the internal rearrangement scenario of
magnetar giant flares, the results of nonlinear simulations can be
used to put an upper limit on the amplitude of the GW signal
potentially emitted in connection to a giant flare and on its level of 
detectability with the next generation GW detectors. 
This idea was put forward in \citet{Ciolfi2011}, where the first
waveform produced in this way was presented. 
The GW emission was then studied systematically in
\citet{Lasky2012,Zink2012,Ciolfi2012}.

The main results on GWs of \citet{Ciolfi2012} are summarized in
Fig.~\ref{gw}, where the (root-sum-square) amplitude of the continuous
emission at the f-mode frequency is plotted against the magnetic field
strength, assuming a typical source distance of 10 kpc and a GW
damping time of 100 ms.  
A good match with the theoretical expectation that the GW amplitude
scales quadratically with the magnetic field strength in the weak
field limit \citep{Levin2011}, allows to safely extrapolate the
amplitude down to the more realistic value of $10^{15}$~G.
By comparing with the sensitivity of the next generation GW detectors, 
the amplitude is found to be orders of magnitude below detectability.  
Since the instability of a purely poloidal field is likely to provide
an upper limit to the realistic GW emission in coincidence with a
giant flare (see Section~\ref{eme}), the prospects of detection
result extremely small in the near future.  
In \citet{Zink2012} and \citet{Lasky2012} the GW amplitudes follow a 
different scaling law ($h\propto B^n$, with $n\sim 3$), but the
prospects of detection are found to be equally pessimistic. 

\section{Concluding remarks}\label{concl}

Understanding the properties of the internal magnetic field of neutron
stars represents one of the most important long-standing open issues
in the physics and astrophysics of these objects.
An important part of the current effort is devoted to the search for
equilibrium configurations of magnetized NSs which are stable on
Alfv\`en timescales, as these would represent a viable description of
the internal magnetic field at the time of crust formation, thus
setting the main properties of the magnetized NS at the beginning of
its long-term dissipative evolution.   
So far, the state-of-the-art nonlinear simulations of magnetized NSs in
general relativity, discussed in this paper, only considered the
simple cases of purely poloidal and purely toroidal magnetic field
geometries, confirming their unstable nature as predicted in previous
linear studies. 
The future of this investigation is to consider mixed
poloidal-toroidal configurations which currently represent good
candidates for stability, as the twisted-torus one. 
In order to improve the realism of the simulations, various aspects
need to be refined, \eg the treatment of magnetic fields in the
magnetosphere surrounding the star. 
Moreover, the present simulations only consider simple polytropic
equations of state and should be extended to more realistic ones. 
Finally, stable stratification due to composition gradients, whose
role in NSs needs to be clarified, could represent an additional
necessary ingredient to be included in the simulations
\citep{Reisenegger2009}.  

As discussed in Section~\ref{GFs}, numerical simulations of the purely
poloidal instability in NSs also provided indications on the
electromagnetic and GW emission to be expected in association with a
hydromagnetic instability, useful as a test case for the internal
rearrangement scenario of magnetar giant flares.
Future work in this direction will help understanding more on this
exciting phenomena.


\acknowledgements

RC is supported by the Humboldt Foundation. Support comes also 
from “CompStar”, a Research Networking Programme of the European
Science Foundation.  



\end{document}